\documentstyle[12pt]{article}

\advance \topmargin by -\headheight
\advance \topmargin by -\headsep

\evensidemargin \oddsidemargin
\begin{document}

\newpage
\pagestyle{empty}

\centerline{\large\bf On the CJT Formalism
in Multi-Field Theories}
\vskip 2 cm
\begin{center}
{\bf Giovanni AMELINO-CAMELIA}\\
\end{center}
\begin{center}
{\it Theoretical Physics, University of Oxford,
1 Keble Rd., Oxford OX1 3NP, UK}
\end{center}
\vskip 1 cm
\centerline{\bf ABSTRACT }
\medskip
The issues that arise when using 
the Cornwall-Jackiw-Tomboulis formalism in multi-field theories
are investigated. 
Particular attention is devoted to the interplay between
temperature effects, ultraviolet structure, and 
the interdependence of the gap equations.
Results are presented explicitly
in the case of the evaluation of the
finite temperature effective potential of a theory
with two scalar fields which has attracted interest
as a toy model for symmetry nonrestoration at high temperatures.
The lowest nontrivial order of approximation
of the Cornwall-Jackiw-Tomboulis effective potential is shown
to lead to consistent results, which are relevant for recent
studies of symmetry nonrestoration by Bimonte and Lozano.

\medskip\noindent{\bf PACS numbers:} 11.30.Qc, 12.10.Dm, 14.80.Hv, 90.80.Cq
\vfill
\noindent{OUTP-96-02-P \space\space\space hep-th/9603135
\hfill January 1996}

\newpage
\pagenumbering{arabic}
\setcounter{page}{1}
\pagestyle{plain}
\baselineskip 12pt plus 0.2pt minus 0.2pt

\section{INTRODUCTION}
$~$

The importance of nonperturbative techniques
in ordinary (``zero-temperature"
or 
``vacuum") 
fied theory
is widely recognized.
These techniques are even more important,
sometimes providing the only consistent approach to a problem,
in finite temperature field theory,
which is affected by infrared problems that are not easily handled
within perturbative approaches.

Over the last twenty years the
Cornwall-Jackiw-Tomboulis (CJT)
formalism of the effective action
for composite operators\cite{corn} has been frequently used
as a nonperturbative technique for the study of zero-temperature
problems. More recently, in Refs.\cite{pap7,pap8}, 
Pi and I advocated the use
of the CJT formalism also in the investigation of problems 
of finite temperature field theory.
Tests\cite{hsugac,quigac} of our proposal have found that it is
among the 
best approaches\footnote{In particular, the approach discussed in
Refs.\cite{linde,dinelinde,hsuboyd} was found to have comparable
qualities.} 
to the study of the type of issues naturally arising in
finite temperature field theory.
In some cases the results of Refs.\cite{pap7,pap8} have been used
as standards to which other nonperturbative techniques
are compared\cite{comp1,comp2}.
These successes of the CJT formalism are however somewhat
limited (especially in the finite temperature context), since
they have been obtained either within rigorous analyses of
single-field\footnote{Notice that the study of the $O(N)$ model 
in the large $N$ approximation, 
as done for example in Ref.\cite{corn}, effectively
reduces the analysis to the one of a 
single-field theory.}
theories\cite{corn,pap7,hsugac,quigac} 
or within analyses of multi-field theories in some rather
drastic approximation\cite{corn,pap8}.
This is a noticeable limitation since the 
CJT formalism can be importantly affected by the presence
of more than one field, requiring the study of
interdependent ultraviolet-divergent 
self-consistent equations.

In order to test the reliability of the approximations based
on the CJT formalism in the context of multi-field theories,
in the present paper I derive the ``bubble approximation''
of the CJT effective potential
in a (thermal) two-scalar-field 
theory, with particular attention to
the interplay between
temperature effects, ultraviolet structure, and 
the interdependence of the gap equations.

Establishing more rigorously the reliability of the CJT formalism 
in multi-field (thermal) theories 
can be very important; for example,
this could be useful
for the investigation
of the possibility of symmetry nonrestoration at high 
temperatures[11-13]
%%%temperatures\cite{wei,mohase,pilang},
a proposal of
great importance for modern cosmology and particle physics,
which has been recently reenergized by the results presented in 
Refs.[14-18].
%%%Refs.\cite{sissa1,sissa2,sissa3,wilsnr,dvalmore}.
Symmetry nonrestoration scenarios usually require
a multi-field theory, and
their investigation has lead to some controversy
(recently revisited in Ref.\cite{sissa3});
in fact, the results on this subject obtained within
certain nonperturbative approximation schemes\cite{nonpertsnr} 
are very different from the ones obtained 
perturbatively[11-18].
%%%perturbatively\cite{wei,mohase,pilang,sissa1,sissa2,sissa3,wilsnr,dvalmore}.
The CJT formalism appears to be ideally suited for the investigation
of this subject, since it encodes some features of
the nonperturbative regime
within a systematic expansion in loops which is
quite similar to those encountered in perturbative approaches;
it might therefore provide the possibility to bridge the conceptual and
quantitative differences 
between perturbative and nonperturbative approaches 
to the study of symmetry nonrestoration.

In the CJT formalism the (thermal)
effective potential $V_T$ is obtained
as the solution of a variational problem for the
effective potential for composite operators $W_T$:
\begin{equation}
V_T(\phi) = W_T[\phi;D_T(\phi;k)]
~,\label{vw}
\end{equation}
\begin{equation}
\biggl[{\delta W_T[\phi;G(k)] \over \delta 
G(k)}\biggr]_{G(k)=D_T(\phi;k)}
= 0
~,\label{hfc}
\end{equation}
where the index $T$ stands for temperature ($T \! = \! 0$
corresponds to the theory in vacuum).
A rigorous definition of $W_T$ can be found 
in Refs.\cite{corn,pap7,pap8,jackbanf}; for the purposes of the
present paper it is sufficient to observe that $W_T$ admits a
loop expansion, with $G(k)$ appearing as the (dressed) propagator:
\begin{equation}
W_T[\phi;G(k)] = V_{tree}(\phi)
+ {1 \over 2} \, \hbox{$\sum$}\!\!\!\!\!\!\!\int^{(T)}_k \, 
Tr [\ln G^{-1}(k) + D_{tree}^{-1}(\phi;k) G(k) -1]
+ W^*_T[\phi;G(k)]
~,\label{hfb}
\end{equation}
where
$W^*_T$ is given by 
all the two-particle-irreducible
vacuum-to-vacuum
graphs with two or more loops
in the theory with vertices given by the interaction part of the 
shifted ($\Phi \rightarrow \Phi + \phi$) Lagrangian and
propagators set equal to $G(k)$.
Also notice that, when $T \! \ne \! 0$, the fourth component of
momentum is discretized, $k_4 = i \pi n T$
($n$ is even for bosons, whereas it is odd for fermions),
as appropriate for the imaginary time formalism of finite temperature
field theory, which I intend to use.
Moreover, in order to be able to discuss at once the
zero-temperature 
and the finite-temperature cases, I introduced the notation
\begin{equation} 
\hbox{$\sum$}\!\!\!\!\!\!\!\int^{(T)}_p ~ \equiv \,
T \sum^{\infty}_{n=-\infty} \int {d^3p \over (2 \pi)^3}
~,
\label{imfeynb}
\end{equation}
which at $T \! = \! 0$ is understood to denote the usual
momentum integration of field theory in vacuum
\begin{equation} 
\hbox{$\sum$}\!\!\!\!\!\!\!\int^{(0)}_p ~ \equiv \,
\int {d^4p \over (2 \pi)^4}
~.
\label{imfeyn}
\end{equation}

The bubble approximation, which I consider in this paper,
is the lowest nontrivial 
order[1-3,20,21]
%%%order\cite{corn,pap7,pap8,jackbanf,consolicjt}
of approximation of the effective
potential in the CJT formalism. It is obtained 
by including in $W^*_T$ 
only the ``double-bubble diagrams'', {\it i.e.}
diagrams with the topology of two rings touching at one point.
Formally, the bubble effective potential can be written as 
\begin{eqnarray}
V^{bubble}_T(\phi) &=& 
W^{bubble}_T[\phi;D^{bubble}_T(\phi;k)] = 
V_{tree}(\phi)
+{1 \over 2} \, \hbox{$\sum$}\!\!\!\!\!\!\!\int^{(T)}_k \, \ln 
[D_T^{bubble}(k)]^{-1}
\nonumber\\
& & 
+ {1 \over 2} \, \hbox{$\sum$}\!\!\!\!\!\!\!\int^{(T)}_k \, 
[D_{tree}^{-1}(\phi;k) D_T^{bubble}(k) -1] 
+ W^{* bubble}_T[\phi;D^{bubble}_T(\phi;k)]
~,\label{hfbubble}
\end{eqnarray}
where $D^{bubble}_T$ is the solution of
\begin{equation}
\biggl[{\delta W^{bubble}_T[\phi;G(k)] \over \delta 
G(k)}\biggr]_{G(k)=D^{bubble}_T(\phi;k)}
= 0
~.\label{hfcbubble}
\end{equation}
A complete discussion of the bubble approximation and some
of its applications of physical relevance are given in
Refs.[1-3,20,21].
%%%Refs.\cite{corn,pap7,pap8,jackbanf,consolicjt}.
For the present paper it is important 
that, for single-field theories, it has been possible
to show the renormalizability and general consistency of
the bubble approximation, and it is therefore
reasonable to test the reliability
of the CJT formalism for multi-field theories
by performing an analogous calculation.
As a preliminary indication of the importance of the
CJT formalism for the study of symmetry nonrestoration,
I also point out that the CJT bubble approximation
leads to the equations on which 
the analysis of Ref.\cite{sissa2} is based.

In order to introduce modularly the various conceptual and technical
issues, the paper is organized as follows.
In the next section, I review the classic zero-temperature
result for 
the single-scalar-field $Z_2$-invariant model, with quartic contact 
interactions.
In Sec.3, I consider 
a two-scalar-field $Z_2 \! \times \! Z_2$-invariant model
at zero temperature, and address the issues introduced
by the interdependence among the corresponding
two gap equations (one for each field) 
of the CJT formalism.
In Sec.4, I finally add thermal effects, and investigate
the resulting structure of the bubble effective potential
in the same two-field model considered in Sec.3.
Sec.5 is devoted to closing remarks.

\section{$Z_2$ MODEL AT $T=0$}
$~$

The single scalar field theory of Euclidean Lagrange density
\begin{equation}
L = {1 \over 2} (\partial_{\mu} \Phi) (\partial^{\mu} \Phi) 
+{1 \over 2} m^2 \Phi^2
+{\lambda_\Phi \over 24}  \Phi^4 ~,
\label{lf}
\end{equation}
has been extensively studied within the bubble approximation
of the CJT formalism.
I review its analysis in order to
set up notation and make observations useful in the study
of the more complex model considered in the later sections.

The interaction Lagrangian 
of the $\Phi \rightarrow \Phi + \phi$ shifted
theory is
\begin{equation}
L_{int}(\phi;\Phi) = {\lambda_\Phi \over 24} \Phi^4 
+ {\lambda_\Phi \over 6} \phi \Phi^3 
~,
\label{sif}
\end{equation}
the tree-level (classical) potential has the form
\begin{equation}
V_{tree} = {m^2 \over 2} \phi^2 + {\lambda_\Phi \over 24} \phi^4
~,
\label{vtree}
\end{equation}
which reflects the $Z_2$ invariance ($\Phi \rightarrow - \Phi$)
of (\ref{lf}),
and the tree-level propagator is
\begin{equation}
D_{tree}(\phi;k) 
= {1 \over k^2 + m^2 + {\lambda_\Phi \over 2} \phi^2 } ~.
\label{dfif}
\end{equation}

From Eqs.(\ref{hfbubble})-(\ref{sif}) 
one finds that the 
zero-temperature CJT bubble\footnote{Notice that in the following,
unlike in the introduction, I suppress the ``bubble" index.} 
potential is given by\cite{pap7}
\begin{eqnarray}
V_0(\phi) &=& {m^2 \over 2} \phi^2 
+ {\lambda_\Phi \over 24} \phi^4
+{1 \over 2} \, \hbox{$\sum$}\!\!\!\!\!\!\!\int^{(0)}_k \, 
\ln D_0^{-1}(\phi;k) 
\nonumber\\
& & + {1 \over 2} \, \hbox{$\sum$}\!\!\!\!\!\!\!\int^{(0)}_k \, 
[(k^2 + m^2 + {\lambda_\Phi \over 2} \phi^2) D_0(\phi;k)  -1] 
+ {\lambda_\Phi \over 8} 
\left[ \hbox{$\sum$}\!\!\!\!\!\!\!\int^{(0)}_k 
D_0(\phi;k) \right]^2
~. \label{gexpy}
\end{eqnarray}
where $D_0(\phi;k)$ is the solution of the bubble-approximated
gap equation (\ref{hfcbubble}), which in the present case can be
written as
\begin{equation}
D_0^{-1}(\phi;k) = 
k^2 + m^2 + {\lambda_\Phi \over 2} \phi^2
+ \, {\lambda_\Phi \over 2} \, 
\hbox{$\sum$}\!\!\!\!\!\!\!\int^{(0)}_p D_0(\phi;p) ~. 
\label{gapgfb}
\end{equation}
The last term on the r.h.s. of Eq.(\ref{gexpy}) 
(which is responsible for the last
term on the r.h.s. of Eq.(\ref{gapgfb}))
is the contribution
of the double-bubble diagram\cite{pap7},
which is the leading 
two-loop contribution to the CJT effective potential
for composite operators in this model.

Without loss of generality one can write
\begin{equation}
D_0(\phi;k)= {1 \over k^2 + M_0^2(\phi;k)}
~,
\label{gansfi}
\end{equation}
and in terms of the ``effective mass" $M_0$ 
the gap equation Eq.(\ref{gapgfb})
can be written as
\begin{equation}
M_0^2(\phi;k) = m^2 + {\lambda_\Phi \over 2} \phi^2 
+ {\lambda_\Phi \over 2} P_0[M_0]
~,
\label{gapth}
\end{equation}
where $P_0[M_0]$ is the zero-temperature limit of
\begin{equation}
P_T[M_T] \, \equiv \, 
\hbox{$\sum$}\!\!\!\!\!\!\!\int^{(T)}_p  
\, {1 \over p^2 + M_T^2(\phi;p)}
~.
\label{gxxth}
\end{equation}
Since $P_0[M_0]$ is momentum independent,
Eq.(\ref{gapth}) implies that (in the bubble approximation)
the effective mass is momentum independent: $M_0 = M_0(\phi)$.

In terms of the solution $M_0(\phi)$ of Eq.(\ref{gapth}), the 
bubble effective potential takes the form
\begin{eqnarray}
V_0(\phi)&=& {m^2 \over 2} \phi^2 + {\lambda_\Phi \over 24} \phi^4
+{1 \over 2} \, \hbox{$\sum$}\!\!\!\!\!\!\!\int^{(0)}_k \, 
\ln [k^2+M_0^2(\phi)] 
\nonumber\\
& & - {1 \over 2} \,  
[M_0^2(\phi)-m^2- {\lambda_\Phi \over 2} \phi^2] \, P_0[M_0(\phi)]
+ {1 \over 8} \lambda_\Phi \left( P_0[M_0(\phi)] \right)^2
\label{vsuma}
\end{eqnarray}

This expression of $V_0$ is affected by two types of divergencies: 
one originating from its divergent integrals,
and the other originating from the fact that
$M_0(\phi)$ is not
well-defined because of the infinities in $P_0(M_0)$. 
Let me start the renormalization procedure by
obtaining a well-defined finite expression for $M_0(\phi)$.
As shown in Ref.\cite{jackprd9}, 
\begin{equation}
P_0[M_0] = 
I_1 - M_0^2 I_2  + {M_0^2 \over 16 \pi^2} 
\ln {M_0^2 \over \mu^2} 
~,
\label{p0reg}
\end{equation}
where $I_{1,2}$ are divergent integrals
\begin{equation}
I_1 \equiv \int {d^3k \over (2 \pi)^3}~ {1 \over 2 |{\bf k}|} =
\lim_{\Lambda \rightarrow \infty} {\Lambda^2 \over 8 \pi^2}
~,
\label{renmlc}
\end{equation}
\begin{equation}
I_2 \equiv \int {d^3k \over (2 \pi)^3}~
[ {1 \over 2 |{\bf k}|} - {1 \over 2 \sqrt{|{\bf k}|^2 + \mu^2}} ] = 
\lim_{\Lambda \rightarrow \infty} {1 \over 16 \pi^2}
\ln {\Lambda^2 \over \mu^2}
~,
\label{renmld}
\end{equation}
$\mu$ is the renormalization scale, and $\Lambda$ is the ultraviolet
momentum cut-off. 

Using Eq.(\ref{p0reg}), the gap equation can be rewritten as
\begin{equation}
M_0^2 = I_1 - M_0^2 I_2 + m^2 + {\lambda_\Phi \over 2} \phi^2 
+ \lambda_\Phi {M_0^2 \over 32 \pi^2} 
\ln {M_0^2 \over \mu^2}
~,
\label{mgapdiv}
\end{equation}
and the divergent terms can be reabsorbed by introducing the following
renormalized parameters ${\widetilde \lambda}_\Phi$
and ${\widetilde m}$ 
\begin{eqnarray}
{1 \over {\widetilde \lambda}_\Phi} &=& {1 \over \lambda_\Phi} +
{I_2 \over 2} 
~,
\label{renmla}\\
{{\widetilde m} \over {\widetilde \lambda}_\Phi} &=& 
{m^2 \over \lambda_\Phi} 
+ {I_1 \over 2} 
~,
\label{renmlb}
\end{eqnarray}
leading to the renormalized gap equation
\begin{equation}
M_0^2 = 
{\widetilde m}^2 + {{\widetilde \lambda}_\Phi \over 2} \phi^2 
+ {\widetilde \lambda}_\Phi {M_0^2 \over 32 \pi^2} 
\ln {M_0^2 \over \mu^2} 
~.
\label{mgapfi}
\end{equation}

Before completing the renormalization of $V_0$, let me discuss
the structure of the renormalized parameters that were just introduced.
In particular, notice that,
in order to keep the renormalized coupling ${\widetilde \lambda}_\Phi$
positive and finite,
the bare coupling must take negatively
vanishing values as the cut-off is  
removed ($\lambda_\Phi \rightarrow 0^-$ as $\Lambda \rightarrow \infty$),
leading to an unstable\cite{mosh} theory.
This is one aspect of the known ``triviality" of the theory under 
consideration; in fact, Eq.(\ref{renmla}) also implies that
the theory becomes free (${\widetilde \lambda}_\Phi \rightarrow 0$)
as the cut-off is removed,
if, as required by stability, the bare coupling is positive.
For physical applications, in which it is desirable to keep
positive both $\lambda_\Phi$ and ${\widetilde \lambda}_\Phi$,
this $Z_2$ model is usually considered as an effective low-energy
theory, with finite cut-off $\Lambda$ such that
\begin{equation}
{{\widetilde \lambda}_\Phi \over 
32 \pi^2} \ln {\Lambda^2 \over \mu^2} < 1
~,
\label{condapi}
\end{equation}
(as required by Eq.(\ref{renmla}) for 
positive $\lambda_\Phi$ and ${\widetilde \lambda}_\Phi$),
but larger than any physical mass scale in the problem
(momenta, temperature, etc.).
Actually, in many applications\cite{pap7} the interesting case 
is
\begin{equation}
{{\widetilde \lambda}_\Phi \over 
32 \pi^2} \ln {\Lambda^2 \over \mu^2} << 1
~,
\label{specialcase}
\end{equation}
which leads to the ideal scenario of cut-off independence\footnote{As
shown in Ref.\cite{pap7}, and reviewed below, 
the cut-off decouples from the analysis
in the limit (\ref{specialcase}).}
with positive $\lambda_\Phi$ and ${\widetilde \lambda}_\Phi$.
Consistently with these observations I am ultimately most interested
in the cases (\ref{condapi})-(\ref{specialcase}), and
I keep track of the ultraviolet cut-off $\Lambda$. 
Renormalizability is obviously encoded in the finiteness of 
the $\Lambda \rightarrow \infty$ limit.

Having clarified these ``triviality-related issues'', 
I can proceed verifying that the 
relations (\ref{renmla})-(\ref{renmlb}), which were
introduced to renormalize the bubble gap equation, also renormalize
the bubble effective potential.
In the simple model presently under consideration this
can be done in several ways\cite{pap7};
I adopt one that can be rather naturally generalized,
as shown in the following sections,
to the case of multi-field theories.
Let me start by noticing that from the known\cite{jackprd9} result
\begin{equation}
\hbox{$\sum$}\!\!\!\!\!\!\!\int^{(0)}_k \, \ln [k^2+M_0^2] \! =
\! {M_0^2 \over 2}  I_1   
- {M_0^4 \over 4}  I_2
+ {M^4_0 \over 64 \pi^2}             
[\ln {M_0^2 \over \mu^2} - {1 \over 2}] 
~,
\label{vone0reg}
\end{equation}
and Eqs.(\ref{gapth}) and (\ref{vsuma}),
it follows that (up to irrelevant $\phi$-independent contributions)
\begin{eqnarray}
V_0 &=&
{m^2 \over 2} \phi^2 + {\lambda_\Phi \over 24} \phi^4
+ {M^4_0 \over 64 \pi^2}             
[\ln {M_0^2 \over \mu^2} - {1 \over 2}] 
- {M_0^4 \over 4}  I_2
+{M_0^2 \over 2}  I_1   \nonumber\\
& &- {1 \over 2 \lambda_\Phi} \,  
[M_0^2 - m^2 - {\lambda_\Phi \over 2} \phi^2]^2 ~ .
\label{vrensuma}
\end{eqnarray}
This can be rewritten using the
definitions (\ref{renmla})-(\ref{renmlb})
as
\begin{eqnarray}
V_0 &=&
{{\widetilde m}^2 \over 2} \phi^2 
+ {{\widetilde \lambda}_\Phi \! - \! {\lambda}_\Phi \over 12} \phi^4
+ {M^4_0 \over 64 \pi^2}             
[\ln {M_0^2 \over \mu^2} - {1 \over 2}] 
\nonumber\\
& &- {1 \over 2 {\widetilde \lambda}_\Phi} \,  
[M_0^2 - {\widetilde m}^2 
- {{\widetilde \lambda}_\Phi \over 2} \phi^2]^2 ~ .
\label{vrensumab}
\end{eqnarray}
Finally,
using the renormalized gap equation,
one finds that 
\begin{eqnarray}
V_0 &=&
{{\widetilde m}^2 \over 2} \phi^2 
+ {{\widetilde \lambda}_\Phi \over 24} \phi^4
+ {{\widetilde \lambda}_\Phi \! - \! {\lambda}_\Phi \over 12} \phi^4
\nonumber\\
& & + {M^4_0 \over 64 \pi^2} [\ln {M_0^2 \over \mu^2} - {1 \over 2}] 
-  {{\widetilde \lambda}_\Phi \over 2} \,  
[{M^2_0 \over 32 \pi^2} \ln {M_0^2 \over \mu^2}]^2 ~,
\label{vzfinal}
\end{eqnarray}
where $M_0$ is the solution of the renormalized
gap equation (\ref{mgapfi}).

The dependence on the cut-off is all included in the
term 
\begin{eqnarray}
{{\widetilde \lambda}_\Phi \! - \! {\lambda}_\Phi \over 12} \phi^4 =
- { {{\widetilde \lambda}_\Phi \over 32 \pi^2} 
\ln {\Lambda^2 \over \mu^2} 
\over
1 - {{\widetilde \lambda}_\Phi \over 32 \pi^2} 
\ln {\Lambda^2 \over \mu^2} }
{{\widetilde \lambda}_\Phi \over 12} \phi^4 ~.
\label{extraterm}
\end{eqnarray}
The renormalizability of the CJT bubble effective potential
of the $Z_2$ model is therefore shown by the fact that the 
$\Lambda \rightarrow \infty$ limit of (\ref{extraterm}) is
well-defined and finite.
The form of the effective potential in the 
limit (\ref{specialcase}) is obtained from (\ref{vzfinal})
by neglecting the term (\ref{extraterm}).

\section{$Z_2 \times Z_2$ MODEL AT $T=0$}
$~$

Still keeping, for the moment, $T \! = \! 0$, I now study
the two-scalar-field theory of Euclidean Lagrange density 
\begin{equation}
L = {1 \over 2} (\partial_{\mu} \Phi) (\partial^{\mu} \Phi) +
{1 \over 2} (\partial_{\mu} \Psi) (\partial^{\mu} \Psi) 
+{1 \over 2} m^2 \Phi^2
+{1 \over 2} \omega^2 \Psi^2
+{\lambda_\Phi \over 24}  \Phi^4
+{\lambda_\Psi \over 24}  \Psi^4
+{\lambda_{\Phi \Psi} \over 4}  \Phi^2 \Psi^2
~,
\label{lfzz}
\end{equation}
which is $Z_2 \times Z_2$ 
invariant [$(\Phi \rightarrow -\Phi) \times (\Psi \rightarrow -\Psi)$].

In general in such a theory one could consider
the effective potential $V(\phi,\psi)$ corresponding to the
shifts $\{ \Phi,\Psi \} \rightarrow \{ \Phi+\phi,\Psi+\psi \}$.
However, for the type of test of the CJT formalism that I am
performing it is sufficient to look at the projection 
of $V(\phi,\psi)$ on the $\psi \! = \! 0$ 
(or equivalently the $\phi \! = \! 0$) axis,
and this is convenient in order to simplify the rather bulky
formulas involved. 
Moreover, 
scenarios for symmetry nonrestoration at high temperatures
within this $Z_2 \times Z_2$ model 
require $\lambda_\Psi \! > \! 
- \lambda_{\Phi \Psi} \! > \! \lambda_\Phi \! > \! 0$
(or, alternatively, $\lambda_\Phi \! > \! 
- \lambda_{\Phi \Psi} \! > \! \lambda_\Psi \! > \! 0$),
in which case
all the significant information is encoded in $V(\phi,\psi \! = \! 0)$
(or, alternatively, $V(\phi \! = \! 0,\psi)$).
Therefore, in the following, I concentrate 
on $V(\phi,\psi \! = \! 0)$, {\it i.e.} 
shifts $\{ \Phi,\Psi \} \rightarrow \{ \Phi+\phi,\Psi \}$,
and, for short, use the notation $V(\phi)$ for (the bubble
approximation of) $V(\phi,\psi \! = \! 0)$.

The shift $\{ \Phi,\Psi \} \rightarrow \{ \Phi+\phi,\Psi \}$
leads to the interaction Lagrangian 
\begin{equation}
L_{int}(\phi;\Phi) = {\lambda_\Phi \over 24} \Phi^4 
+ {\lambda_\Psi \over 24} \Psi^4 
+ {\lambda_{\Phi} \over 6} \phi \Phi^3 
+ {\lambda_{\Phi \Psi} \over 2} \phi \Phi^3 
~,
\label{sifzz}
\end{equation}
the tree-level potential
\begin{equation}
V_{tree} = {m^2 \over 2} \phi^2 + {\lambda_\Phi \over 24} \phi^4
~,
\label{vtreezz}
\end{equation}
and the tree-level propagator
\begin{equation}
[D_{tree}(\phi;k)]_{ab}
= {\delta_{a1} \delta_{b1} 
\over k^2 + m^2 + {\lambda_\Phi \over 2} \phi^2 } +
{\delta_{a2} \delta_{b2} 
\over k^2 + \omega^2 + {\lambda_{\Phi \Psi} \over 2} \phi^2 } 
~.
\label{dfifzz}
\end{equation}

The zero-temperature bubble effective potential is given by
\begin{eqnarray}
V_0 \!\!&=&\!\! {m^2 \over 2} \phi^2 
+ {\lambda_\Phi \over 24} \phi^4
- {1 \over 2} \, \hbox{$\sum$}\!\!\!\!\!\!\!\int^{(0)}_k \, 
\{ \ln [D_0(\phi;k)]_{11} 
+ \ln [D_0(\phi;k)]_{22} \}
\nonumber\\
& & 
+ {1 \over 2} \, \hbox{$\sum$}\!\!\!\!\!\!\!\int^{(0)}_k \, 
\{ (k^2 \! + \! m^2 \! + \! {\lambda_\Phi \over 2} \phi^2 )
[D_0(\phi;k)]_{11} 
+ (k^2 \! + \! \omega^2 \! 
+ \! {\lambda_{\Phi \Psi} \over 2} \phi^2) 
[D_0(\phi;k)]_{22} -2 \} \nonumber\\
& & 
+ {\lambda_\Phi \over 8} 
\left[ \hbox{$\sum$}\!\!\!\!\!\!\!\int^{(0)}_k 
\, [D_0(\phi;k)]_{11} \right]^2
+ {\lambda_\Psi \over 8} 
\left[ \hbox{$\sum$}\!\!\!\!\!\!\!\int^{(0)}_k 
\, [D_0(\phi;k)]_{22} \right]^2
\nonumber\\
& & 
+ {\lambda_{\Phi \Psi} \over 4} 
\hbox{$\sum$}\!\!\!\!\!\!\!\int^{(0)}_k 
\, [D_0(\phi;k)]_{11} 
\hbox{$\sum$}\!\!\!\!\!\!\!\int^{(0)}_p
\, [D_0(\phi;p)]_{22} 
~. \label{gexpyzz}
\end{eqnarray}
where $[D_0(\phi;k)]_{11}$ and $[D_0(\phi;k)]_{22}$
are the solutions of the gap equations
\begin{eqnarray}
([D_0(\phi;k)]_{11})^{-1}
\! &=& \! ([D_{tree}(\phi;k)]_{11})^{-1}
+ \, {\lambda_\Phi \over 2} 
\, \hbox{$\sum$}\!\!\!\!\!\!\!\int^{(0)}_p 
\, [D_0(p)]_{11}
+ \, {\lambda_{\Psi \Psi} \over 2} 
\, \hbox{$\sum$}\!\!\!\!\!\!\!\int^{(0)}_p 
\, [D_0(p)]_{22} ~, \nonumber\\
([D_0(\phi;k)]_{22})^{-1} \! &=& \! ([D_{tree}(\phi;k)]_{22})^{-1}
+ \, {\lambda_\Psi \over 2} 
\, \hbox{$\sum$}\!\!\!\!\!\!\!\int^{(0)}_p 
\, [D_0(p)]_{22} 
+ \, {\lambda_{\Psi \Psi} \over 2} 
\, \hbox{$\sum$}\!\!\!\!\!\!\!\int^{(0)}_p 
\, [D_0(p)]_{11} 
~. 
\label{gapgfbzz}
\end{eqnarray}

Again, it is convenient to reexpress the effective propagator $D_0$
in terms of effective masses
\begin{equation}
[D_0(\phi;k)]_{ab} = {\delta_{a1} \delta_{b1} 
 \over k^2 + M_0^2(\phi;k)} +
{\delta_{a2} \delta_{b2} \over k^2 + \Omega_0^2(\phi;k)}
~,
\label{gansfizz}
\end{equation}
allowing to rewrite the gap equations as
\begin{eqnarray}
M_0^2(\phi;k) &=& m^2 + {\lambda_\Phi \over 2} \phi^2 
+ {\lambda_\Phi \over 2} P_0[M_0] 
+ {\lambda_{\Phi \Psi} \over 2} P_0[\Omega_0]
~,
\nonumber\\
\Omega_0^2(\phi;k) &=& \omega^2 
+ {\lambda_{\Phi \Psi} \over 2} \phi^2 
+ {\lambda_\Psi \over 2} P_0[\Omega_0]
+ {\lambda_{\Phi \Psi} \over 2} P_0[M_0] 
~.
\label{gapthzz}
\end{eqnarray}
This shows that also
in this two-field theory
the effective masses are momentum independent 
within the bubble approximation:
$M_0=M_0(\phi)$, $\Omega_0 = \Omega_0(\phi)$.

In terms of effective masses and bare parameters,
$V_0$ has the form
\begin{eqnarray}
V_0&=& {m^2 \over 2} \phi^2 
+ {\lambda_\Phi \over 24} \phi^4
+{1 \over 2} \, 
\hbox{$\sum$}\!\!\!\!\!\!\!\int^{(0)}_k \, 
\{ \ln [k^2+M_0^2(\phi)] + \ln [k^2+\Omega_0^2(\phi)] \}
\nonumber\\
& & - {1 \over 2} \,  
[M_0^2(\phi) \! - \! m^2 \! 
- \! {\lambda_\Phi \over 2} \phi^2] \, P_0[M_0]
- {1 \over 2} \,  
[\Omega_0^2(\phi) \! - \! \omega^2 \! 
- \! {\lambda_{\Phi \Psi} \over 2} \phi^2] \, 
P_0[\Omega_0]
\nonumber\\
& & + {\lambda_\Phi  \over 8} 
\left( P_0[M_0] \right)^2
+ {\lambda_\Psi \over 8} 
\left( P_0[\Omega_0] \right)^2
+ {\lambda_{\Phi \Psi}  \over 4} 
P_0[M_0] P_0[\Omega_0]
~.
\label{vsumazz}
\end{eqnarray}
The first step toward the renormalization of $V_0$
is the renormalization of the gap equations, 
which, using Eq.(\ref{p0reg}),
can be rewritten as
\begin{eqnarray}
M_0^2(\phi;k) &=& m^2 + {\lambda \over 2} \phi^2 
+ {\lambda_\Phi \over 2} 
\left( I_1 - M_0^2 I_2  + {M_0^2 \over 16 \pi^2} 
\ln {M_0^2 \over \mu^2} \right)
\nonumber\\
& & + {\lambda_{\Phi \Psi} \over 2} 
\left( I_1 - \Omega_0^2 I_2  + {\Omega_0^2 \over 16 \pi^2} 
\ln {\Omega_0^2 \over \mu^2} \right)
~,
\nonumber\\
\Omega_0^2(\phi;k) &=& \omega^2 + {\lambda_{\Phi \Psi} \over 2} \phi^2 
+ {\lambda_\Psi \over 2} 
\left( I_1 - \Omega_0^2 I_2  + {\Omega_0^2 \over 16 \pi^2} 
\ln {\Omega_0^2 \over \mu^2} \right)
\nonumber\\
& & + {\lambda_{\Phi \Psi} \over 2} 
\left( I_1 - M_0^2 I_2  + {M_0^2 \over 16 \pi^2} 
\ln {M_0^2 \over \mu^2} \right)
~.
\label{gapdivzz}
\end{eqnarray}
Notice that the interdependence of the gap equations affects importantly
the structure of divergent terms.
Since in each gap equation divergent coefficients appear 
in front of both $M_0^2$ and $\Omega_0^2$, in this two-field theory
the renormalization cannot proceed by considering
the gap equations independently (whereas 
the only gap equation present in the single-field
theory considered earlier could obviously be renormalized on its own).
Nevertheless, I am able to obtain renormalized 
results by exploiting
the fact that combining appropriately the gap Eqs.(\ref{gapdivzz})
one can derive the following equivalent set of equations
\begin{eqnarray}
0 &=& {\phi^2 \over 2} 
+ {M_0^2 \over 32 \pi^2} \ln {M_0^2 \over \mu^2}
+ {I_1 \over 2} 
+ {\lambda_\Psi m^2  
- \lambda_{\Phi \Psi} \omega^2  
\over \lambda_\Phi \lambda_\Psi 
- \lambda_{\Phi \Psi}^2} 
\nonumber\\
& & 
- \left( {I_2 \over 2} 
+ {\lambda_\Psi \over \lambda_\Phi \lambda_\Psi 
- \lambda_{\Phi \Psi}^2} \right) M_0^2  
+ {\lambda_{\Phi \Psi} \over \lambda_\Phi \lambda_\Psi 
- \lambda_{\Phi \Psi}^2} \Omega_0^2  
~,
\nonumber\\
0 &=& {\Omega_0^2 \over 32 \pi^2} \ln {\Omega_0^2 \over \mu^2}
+ {I_1 \over 2} 
+ {\lambda_\Phi \omega^2  
- \lambda_{\Phi \Psi} m^2  \over \lambda_\Phi \lambda_\Psi 
- \lambda_{\Phi \Psi}^2} 
\nonumber\\
& & 
- \left( {I_2 \over 2} 
+ {\lambda_\Phi \over \lambda_\Phi \lambda_\Psi 
- \lambda_{\Phi \Psi}^2} \right) \Omega_0^2  
+ {\lambda_{\Phi \Psi} \over \lambda_\Phi \lambda_\Psi 
- \lambda_{\Phi \Psi}^2} M_0^2  
~.
\label{gapdivzznew}
\end{eqnarray}
Notice that in the first (second) of these equations
divergent coefficients appear only
in front of $M_0$ ($\Omega_0$).
The structure of the Eqs.(\ref{gapdivzznew}) suggests the introduction
of renormalized 
parameters ${\widetilde \lambda_\Phi}$,
${\widetilde \lambda_\Psi}$,
${\widetilde \lambda_{\Phi \Psi}}$,
${\widetilde m}$, ${\widetilde \omega}$, 
defined by 
\begin{eqnarray}
{{\widetilde \lambda}_\Psi \over 
{\widetilde \lambda}_\Phi {\widetilde \lambda}_\Psi 
- {\widetilde \lambda}_{\Phi \Psi}^2} &=&
{I_2 \over 2}  
+ {\lambda_\Psi \over \lambda_\Phi \lambda_\Psi 
- \lambda_{\Phi \Psi}^2}
~,
\label{renmlzza}\\
{{\widetilde \lambda}_\Phi \over 
{\widetilde \lambda}_\Phi {\widetilde \lambda}_\Psi 
- {\widetilde \lambda}_{\Phi \Psi}^2} &=&
{I_2 \over 2}  
+ {\lambda_\Phi \over \lambda_\Phi \lambda_\Psi 
- \lambda_{\Phi \Psi}^2}
~,
\label{renmlzzb}\\
{{\widetilde \lambda}_{\Phi \Psi} \over {\widetilde \lambda}_\Phi 
{\widetilde \lambda}_\Psi 
- {\widetilde \lambda}_{\Phi \Psi}^2} &=&
{\lambda_{\Phi \Psi} \over \lambda_\Phi \lambda_\Psi 
- \lambda_{\Phi \Psi}^2}
~,
\label{renmlzzc}\\
{{\widetilde \lambda}_\Phi {\widetilde \omega}^2  
- {\widetilde \lambda}_{\Phi \Psi} {\widetilde m}^2  \over 
{\widetilde \lambda}_\Phi {\widetilde \lambda}_\Psi 
- {\widetilde \lambda}_{\Phi \Psi}^2} 
&=&{I_1 \over 2} 
+ {\lambda_\Psi m^2  
- \lambda_{\Phi \Psi} \omega^2 \over \lambda_\Phi \lambda_\Psi 
- \lambda_{\Phi \Psi}^2} 
~,
\label{renmlzzd}\\
{{\widetilde \lambda}_\Psi {\widetilde m}^2  
- {\widetilde \lambda}_{\Phi \Psi} {\widetilde \omega}^2  
\over {\widetilde \lambda}_\Phi 
{\widetilde \lambda}\Psi 
- {\widetilde \lambda}_{\Phi \Psi}^2} 
&=&{I_1 \over 2} 
+ {\lambda_\Psi m^2  
- \lambda_{\Phi \Psi} \omega^2  \over \lambda_\Phi \lambda_\Psi 
- \lambda_{\Phi \Psi}^2} 
~,
\label{renmlzze}
\end{eqnarray}

In terms of these renormalized parameters the Eqs.(\ref{gapdivzznew})
can be rewritten as
\begin{eqnarray}
0 &=& {\phi^2 \over 2} 
+ {M_0^2 \over 32 \pi^2} \ln {M_0^2 \over \mu^2}
+ {{\widetilde \lambda}_\Psi {\widetilde m}^2  
- {\widetilde \lambda}_{\Phi \Psi} {\widetilde \omega}^2  
- {\widetilde \lambda}_\Psi M_0^2  
+ {\widetilde \lambda}_{\Phi \Psi} \Omega_0^2  
\over {\widetilde \lambda}_\Phi 
{\widetilde \lambda}_\Psi 
- {\widetilde \lambda}_{\Phi \Psi}^2} 
~,
\nonumber\\
0 &=& {\Omega_0^2 \over 32 \pi^2} \ln {\Omega_0^2 \over \mu^2}
+ {{\widetilde \lambda}_\Phi {\widetilde \omega}^2  
- {\widetilde \lambda}_{\Phi \Psi} {\widetilde m}^2  
- {\widetilde \lambda}_\Phi \Omega_0^2  
+ {\widetilde \lambda}_{\Phi \Psi} M_0^2  
\over {\widetilde \lambda}_\Phi 
{\widetilde \lambda}\Psi 
- {\widetilde \lambda}_{\Phi \Psi}^2} 
~.
\label{gapfinzznew}
\end{eqnarray}
In turn these equations can be combined to obtain
the following equivalent set of equations
\begin{eqnarray}
M_0^2(\phi;k) &=& {\widetilde m}^2 
+ {{\widetilde \lambda}_\Phi \over 2} \phi^2 
+ {{\widetilde \lambda}_\Phi \over 2} 
{M_0^2 \over 16 \pi^2} 
\ln {M_0^2 \over \mu^2} 
+ {{\widetilde \lambda}_{\Phi \Psi} \over 2} 
{\Omega_0^2 \over 16 \pi^2} 
\ln {\Omega_0^2 \over \mu^2} 
~,
\nonumber\\
\Omega_0^2(\phi;k) &=& {\widetilde \omega}^2 
+ {{\widetilde \lambda}_{\Phi \Psi} \over 2} \phi^2 
+ {{\widetilde \lambda}_\Psi \over 2} 
{\Omega_0^2 \over 16 \pi^2} 
\ln {\Omega_0^2 \over \mu^2} 
+ {{\widetilde \lambda}_{\Phi \Psi} \over 2}  {M_0^2 \over 16 \pi^2} 
\ln {M_0^2 \over \mu^2} 
~,
\label{gapfinzz}
\end{eqnarray}
which can be interpreted as renormalized gap equations.

Before completing the renormalization of $V_0$,
I want to observe that
also the $Z_2 \times Z_2$ model presently under consideration
presents triviality-related features.
In fact, 
from (\ref{renmlzza})-(\ref{renmlzze}) it follows that by requiring
\begin{eqnarray}
{\widetilde \lambda}_\Psi > 0 ~,~~~
{\widetilde \lambda}_\Phi > 0 ~,~~~
{\widetilde \lambda}_\Psi {\widetilde \lambda}_\Phi
- {\widetilde \lambda}_{\Phi \Psi}^2 > 0
~
\label{cond}
\end{eqnarray}
(the conditions usually assumed\cite{wei,sissa2} to be sufficient to 
ensure the stability of the
theory), one finds that
$\lambda_\Psi \rightarrow 0^-$ and $\lambda_\Phi \rightarrow 0^-$ 
in the $\Lambda \rightarrow \infty$ limit,
whereas by insisting on a bare potential bounded from below,
{\it i.e.} demanding
\begin{eqnarray}
\lambda_\Psi > 0 ~,~~~
\lambda_\Phi > 0 ~,~~~
\lambda_\Psi \lambda_\Phi
- \lambda_{\Phi \Psi}^2 > 0 
~,
\label{condbare}
\end{eqnarray}
one finds that ${\widetilde \lambda}_\Psi \rightarrow 0^+$ 
and ${\widetilde \lambda}_\Phi \rightarrow 0^+$ 
in the $\Lambda \rightarrow \infty$ limit.

\noindent
A physically meaningful $Z_2 \times Z_2$ model
can be certainly obtained as a low-energy effective theory
with cut-off $\Lambda$ small enough to be consistent with 
both (\ref{cond})
and (\ref{condbare}), so, like in the previous section,
I keep track of the ultraviolet cut-off $\Lambda$, 
and take the limit $\Lambda \rightarrow \infty$ only when
testing renormalizability.

In order to show that the renormalization 
prescriptions (\ref{renmlzza})-(\ref{renmlzze})
also renormalize the effective potential,
I rewrite $V_0$ using Eqs.(\ref{vone0reg}),
(\ref{vsumazz}), and (\ref{gapdivzznew})
\begin{eqnarray}
V_0 \!\!\! &=& \!\!\!
{m^2 \over 2} \phi^2  +  {\lambda \over 24} \phi^4
- {M_0^4 + \Omega_0^4 \over 4}  I_2 
+ {M_0^2 + \Omega_0^2 \over 2}  I_1   
+ {M^4_0 \over 64 \pi^2}             
[\ln \! {M_0^2 \over \mu^2}  -  {1 \over 2}]  
+ {\Omega^4_0 \over 64 \pi^2}             
[\ln \! {\Omega_0^2 \over \mu^2}  -  {1 \over 2}] 
\nonumber\\
& & 
\!\!\!\!\!\!\!\!\!\!\! \!\!\!- {\lambda_\Phi \over 2}
\left[ { \lambda_{\Phi \Psi} (\Omega_0^2 - \omega^2)
- \lambda_\Psi (M_0^2 - m^2)
+ (\lambda_\Phi \lambda_\Psi - \lambda_{\Phi \Psi}) {\phi^2 \over 2}
\over \lambda_\Phi \lambda_\Psi - \lambda_{\Phi \Psi}^2 } \right]^2
\nonumber\\
& & 
\!\!\!\!\!\!\!\!\!\!\! \!\!\!- {\lambda_\Psi \over 2}
\left[ { \lambda_{\Phi \Psi} (M_0^2 - m^2)
- \lambda_\Phi (\Omega_0^2 - \omega^2)
\over \lambda_\Phi \lambda_\Psi - \lambda_{\Phi \Psi}^2 } \right]^2
\nonumber\\
& & 
\!\!\!\!\!\!\!\!\!\!\! \!\!\!
- \lambda_{\Phi \Psi} \! { [ \lambda_{\Phi \Psi} 
(\Omega_0^2 \!\! - \!\! \omega^2)  \!
- \! \lambda_\Psi (M_0^2 \!\! - \!\! m^2)
\! + \! (\lambda_\Phi \lambda_\Psi \! 
- \! \lambda_{\Phi \Psi}) {\phi^2 \over 2}]
[ \lambda_{\Phi \Psi} (M_0^2 \!\! - \!\! m^2) \!
- \! \lambda_\Phi (\Omega_0^2 \!\! - \!\! \omega^2)]
\over [\lambda_\Phi \lambda_\Psi \! - \! \lambda_{\Phi \Psi}^2 ]^2 }
~. \label{vrensumazz}
\end{eqnarray}
Using the definitions (\ref{renmlzza})-(\ref{renmlzze}) one then
finds that
\begin{eqnarray}
V_0 \!\! \! &=& \!\! \! 
{{\widetilde m}^2 \over 2} \phi^2 +
{{\widetilde \lambda}_\Phi \over 24} \phi^4
+ {{\widetilde \lambda}_\Phi \! - \! {\lambda}_\Phi \over 12} \phi^4
+ {M^4_0 \over 64 \pi^2}             
[\ln {M_0^2 \over \mu^2} - {1 \over 2}] 
+ {\Omega^4_0 \over 64 \pi^2}             
[\ln {\Omega_0^2 \over \mu^2} - {1 \over 2}] ~~~~~~~~~~~~~~~~
\nonumber\\
& & 
\!\!\!\!\!\!\!\!\!\!\! 
- {{\widetilde \lambda}_\Phi \over 2}
\left[ { {\widetilde \lambda}_{\Phi \Psi} (\Omega_0^2 - \omega^2)
- {\widetilde \lambda}_\Psi (M_0^2 - m^2)
+ ({\widetilde \lambda}_\Phi {\widetilde \lambda}_\Psi 
- {\widetilde \lambda}_{\Phi \Psi}) {\phi^2 \over 2}
\over {\widetilde \lambda}_\Phi {\widetilde \lambda}_\Psi 
- {\widetilde \lambda}_{\Phi \Psi}^2 } \right]^2
\nonumber\\
& & \!\!\!\!\!\!\!\!\!\!\! 
- {{\widetilde \lambda}_\Psi \over 2}
\left[ { {\widetilde \lambda}_{\Phi \Psi} (M_0^2 - m^2)
- {\widetilde \lambda}_\Phi (\Omega_0^2 - \omega^2)
\over {\widetilde \lambda}_\Phi {\widetilde \lambda}_\Psi 
- {\widetilde \lambda}_{\Phi \Psi}^2 } \right]^2
\nonumber\\
& & \!\!\!\!\!\!\!\!\!\!\! 
- {\widetilde \lambda}_{\Phi \Psi} \!
{ [{\widetilde \lambda}_{\Phi \Psi} 
(\Omega_0^2  \! \! -  \! \! \omega^2) \!
- {\widetilde \lambda}_\Psi (M_0^2 \!\! - \!\! m^2)
\! + \! ({\widetilde \lambda}_\Phi {\widetilde \lambda}_\Psi 
\! - \! {\widetilde \lambda}_{\Phi \Psi}) {\phi^2 \over 2}]
[ {\widetilde \lambda}_{\Phi \Psi} (M_0^2  \! \! -  \! \! m^2) \!
- \! {\widetilde \lambda}_\Phi (\Omega_0^2 \! \! - \! \! \omega^2)]
\over {\widetilde \lambda}_\Phi {\widetilde \lambda}_\Psi 
- {\widetilde \lambda}_{\Phi \Psi}^2 }
~, \label{vrensumazzren}
\end{eqnarray}
which using Eqs.(\ref{gapfinzznew}) finally leads to the result
\begin{eqnarray}
V_0 \! \! &=& \! \! 
{{\widetilde m}^2 \over 2} \phi^2 +
{{\widetilde \lambda}_\Phi \over 24} \phi^4
+ {{\widetilde \lambda}_\Phi \! - \! {\lambda}_\Phi \over 12} \phi^4
+ {M^4_0 \over 64 \pi^2}             
[\ln {M_0^2 \over \mu^2} - {1 \over 2}] 
+ {\Omega^4_0 \over 64 \pi^2}             
[\ln {\Omega_0^2 \over \mu^2} - {1 \over 2}] 
\nonumber\\
& & - {{\widetilde \lambda}_\Phi \over 8} 
[{M^2_0 \over 16 \pi^2} \ln {M_0^2 \over \mu^2}]^2
- {{\widetilde \lambda}_\Psi \over 8} 
[{\Omega^2_0 \over 16 \pi^2} \ln {\Omega_0^2 \over \mu^2}]^2
- {{\widetilde \lambda}_{\Phi \Psi} \over 4} 
[{M^2_0 \over 16 \pi^2} \ln {M_0^2 \over \mu^2}]
[{\Omega^2_0 \over 16 \pi^2} \ln {\Omega_0^2 \over \mu^2}]
~. \label{vrensumazzrenfinal}
\end{eqnarray}
Again, the only $\Lambda$-dependent contribution
comes from the 
term $({\widetilde \lambda}_\Phi \! - \! {\lambda}_\Phi) \phi^4 /12$.
As it can be easily derived from Eqs.(\ref{renmlzza})-(\ref{renmlzze}),
this term has 
a well-defined and finite $\Lambda \rightarrow \infty$ limit, 
indicating that the CJT bubble potential
of the $Z_2 \times Z_2$ model at zero temperature is renormalizable.

\section{$Z_2 \times Z_2$ MODEL AT FINITE $T$}
$~$

In this section I finally consider the finite temperature case;
specifically, I study the CJT bubble effective potential of 
the $Z_2 \times Z_2$ model in the imaginary time formalism 
of finite temperature field theory.
The temperature dependence, besides introducing additional elements
of technical difficulty, affects very importantly some of the issues
under investigation in the present paper.
In ordinary analyses of perturbative renomarlizability of the finite 
temperature effective potential a central role is played
by the fact the renormalization prescriptions for the parameters
of a field theory are temperature independent\cite{jackbanf,kapu}.
However, as shown below, because of the temperature dependence
of the effective propagator (which results from the nonperturbative 
nature of the approach) there are finite temperature Feynman
diagrams relevant for the CJT formalism that give 
highly nontrivial temperature- and $\phi$- dependent 
divergent contributions
to the effective potential.
Unless these divergencies can be reabsorbed by the introduction
of temperature-independent renormalized parameters,
the physical consistency of the nonperturbative approach is
to be doubted\cite{jackbanf}.

I start by introducing temperature-dependent effective masses
\begin{equation}
[D_T(\phi;k)]_{ab} = {\delta_{a1} \delta_{b1} 
 \over k^2 + M_T^2(\phi;k)} +
{\delta_{a2} \delta_{b2} \over k^2 + \Omega_T^2(\phi;k)}
~,
\label{gansfizzt}
\end{equation}
which, in the bubble approximation, must satisfy the gap equations
\begin{eqnarray}
M_T^2(\phi;k) &=& m^2 + {\lambda_\Phi \over 2} \phi^2 
+ {\lambda_\Phi \over 2} P_T[M_T] 
+ {\lambda_{\Phi \Psi} \over 2} P_T[\Omega_T]
~,
\nonumber\\
\Omega_T^2(\phi;k) &=& \omega^2 
+ {\lambda_{\Phi \Psi} \over 2} \phi^2 
+ {\lambda_\Psi \over 2} P_T[\Omega_T]
+ {\lambda_{\Phi \Psi} \over 2} P_T[M_T] 
~,
\label{gapthzzt}
\end{eqnarray}
indicating that $M_T$ and $\Omega_T$ are momentum 
independent: $M_T \! = \! M_T(\phi)$, $\Omega_T \! = \! \Omega_T(\phi)$.

In terms of the 
effective masses the bubble
effective potential can be written as
\begin{eqnarray}
V_T &=&
{{m}^2 \over 2} \phi^2 +
{{\lambda}_\Phi \over 24} \phi^4
+{1 \over 2} \, 
\hbox{$\sum$}\!\!\!\!\!\!\!\int^{(T)}_k \, 
\{ \ln [k^2+M_T^2] + \ln [k^2+\Omega_T^2] \}
\nonumber\\
& & - {1 \over 2} \,  
[M_T^2 \! - \! m^2 \! 
- \! {\lambda_\Phi \over 2} \phi^2] \, P_T[M_T]
- {1 \over 2} \,  
[\Omega_T^2 \! - \! \omega^2 \! 
- \! {\lambda_{\Phi \Psi} \over 2} \phi^2] \, 
P_T[\Omega_T]
\nonumber\\
& & + {\lambda_\Phi  \over 8} 
\left( P_T[M_T] \right)^2
+ {\lambda_\Psi \over 8} 
\left( P_T[\Omega_T] \right)^2
+ {\lambda_{\Phi \Psi}  \over 4} 
P_T[M_T] P_T[\Omega_T]
\nonumber\\
&=& {{m}^2 \over 2} \phi^2 +
{{\lambda}_\Phi \over 24} \phi^4
+{1 \over 2} \, 
\hbox{$\sum$}\!\!\!\!\!\!\!\int^{(T)}_k \, 
\{ \ln [k^2+M_T^2] + \ln [k^2+\Omega_T^2] \}
\nonumber\\
& & - {\lambda_\Phi  \over 8} 
\left( P_T[M_T] \right)^2
- {\lambda_\Psi \over 8} 
\left( P_T[\Omega_T] \right)^2
- {\lambda_{\Phi \Psi}  \over 4} 
P_T[M_T] P_T[\Omega_T]
~,
\label{vsumazzt}
\end{eqnarray}
where the last equality follows from the gap 
equations (\ref{gapthzzt}).

The ultraviolet divergent contributions can be identified using the
well-known results\cite{doja}
for the ``tadpole" 
\begin{eqnarray}
P_T[M] & = & I_1 - M^2 I_2 + P_T^{(f)}[M]
\nonumber\\
P_T^{(f)}[M] & \equiv & {M^2 \over 16 \pi^2} 
\ln {M^2 \over \mu^2} 
- \int {d^3k \over (2 \pi)^3}~
\left[\sqrt{|{\bf k}|^2 + M^2}  
\left( 1 - exp \left( { \sqrt{|{\bf k}|^2 
+ M^2} \over T} \right) \right) \right]^{-1}
~,
\label{gxxb}
\end{eqnarray}
and the ``one loop''
\begin{eqnarray}
\hbox{$\sum$}\!\!\!\!\!\!\!\int^{(T)}_k \, 
\ln [k^2+M^2] \!\! & = & \!\! 
- {M^4 \over 4}  I_2
+{M^2 \over 2}  I_1 + Q_T^{(f)}[M]
\nonumber\\
Q_T^{(f)}[M] \!\! & \equiv & \!\! {M^4 \over 64 \pi^2}             
[\ln {M^2 \over \mu^2} - {1 \over 2}] 
+ T \int {d^3k \over (2 \pi)^3}~
\ln \left[ 1- 
exp \left( { \sqrt{|{\bf k}|^2 + M^2} \over T} \right) \right]
~.
\label{voneTreg}
\end{eqnarray}
$P_T^{(f)}$ and $Q_T^{(f)}$ are finite ({\it i.e.} do not diverge as the
cut-off is removed) and 
their structure is known\cite{doja} within a ``high temperature''
(small $M/T$) expansion
\begin{eqnarray}
P_T^{(f)}[M] \!\! &\simeq& \!\! 
{T^2 \over 12} - {M T \over 4 \pi}
+ {M^2 \over 16 \pi^2} \ln {M^2 \over T^2} 
\nonumber\\
Q_T^{(f)}[M] \!\! &\simeq& \!\! - {\pi^2 T^4 \over 90} 
+ {M^2 T^2 \over 24} - {M^3 T \over 12 \pi}
- {M^4 \over 64 \pi^2} \ln {M^2 \over T^2} 
~.
\label{pqexp}
\end{eqnarray}

Eqs.(\ref{vsumazzt})-(\ref{pqexp})
show that some of
the divergent contributions to $V_T$
come from terms that are analogous to those encountered in the 
zero-temperature case, but now involve the temperature dependent
effective masses $M_T$,$\Omega_T$ in place of their
zero-temperature limits $M_0$,$\Omega_0$.
Other divergent contributions to $V_T$
come from products of two $P_T$'s,
and involve terms of structure $P_T^{(f)} I_1$ or $P_T^{(f)} I_2$.
The analysis of this latter type of divergent contributions 
is complicated by the fact that
the information available on $P_T^{(f)}$
is only in the form of a series expansion.
However, I show below that 
the steps for renormalization that I followed 
in the previous sections can be generalized to the finite 
temperature case in such a way to require no explicit
information on $P_T^{(f)}$
(besides the fact that it stays finite as the
cut-off is removed).

First, let me reexpress the 
gap equations (\ref{gapthzzt})
in the spirit of the Eqs.(\ref{gapdivzznew}):
\begin{eqnarray}
0 &=& {\phi^2 \over 2} + P^{(f)}_T (M_T)
+ {I_1 \over 2} 
+ {\lambda_\Psi m^2  
- \lambda_{\Phi \Psi} \omega^2  \over \lambda_\Phi \lambda_\Psi 
- \lambda_{\Phi \Psi}^2} 
\nonumber\\
& & 
- \left( {I_2 \over 2} 
- {\lambda_\Psi \over \lambda_\Phi \lambda_\Psi 
- \lambda_{\Phi \Psi}^2} \right) M_T^2  
+ {\lambda_{\Phi \Psi} \over \lambda_\Phi \lambda_\Psi 
- \lambda_{\Phi \Psi}^2} \Omega_T^2  
~,
\nonumber\\
0 &=& P^{(f)}_T (\Omega_T) + {I_1 \over 2} 
+ {\lambda_\Phi \omega^2  - \lambda_{\Phi \Psi} m^2  
\over \lambda_\Phi \lambda_\Psi 
- \lambda_{\Phi \Psi}^2} 
\nonumber\\
& & 
- \left( {I_2 \over 2} 
- {\lambda_\Phi \over \lambda_\Phi \lambda_\Psi 
- \lambda_{\Phi \Psi}^2} \right) \Omega_T^2  
+ {\lambda_{\Phi \Psi} \over \lambda_\Phi \lambda_\Psi 
- \lambda_{\Phi \Psi}^2} M_T^2  
~.
\label{gapdivzztnew}
\end{eqnarray}
In agreement with the physical arguments\cite{jackbanf,kapu}
indicating that the renormalization prescriptions for the
parameters of a thermal field theory should 
be temperature-independent, the divergencies in
these equations can be absorbed by
introducing the 
same renormalized parameters 
found necessary in the 
zero-temperature analysis;
in fact, in terms of the
${\widetilde m}$, ${\widetilde \omega}$, 
${\widetilde \lambda_\Phi}$, ${\widetilde \lambda_\Psi}$,
${\widetilde \lambda_{\Phi \Psi}}$ 
of Eqs.(\ref{renmlzza})-(\ref{renmlzze}),
the Eqs.(\ref{gapdivzztnew}) can be rewritten as
\begin{eqnarray}
0 &=& {\phi^2 \over 2} + P^{(f)}_T (M_T) 
+ {{\widetilde \lambda}_\Psi {\widetilde m}^2  
- {\widetilde \lambda}_{\Phi \Psi} {\widetilde \omega}^2  
- {\widetilde \lambda}_\Psi M_T^2  
+ {\widetilde \lambda}_{\Phi \Psi} \Omega_T^2  
\over {\widetilde \lambda}_\Phi {\widetilde \lambda}_\Psi 
- {\widetilde \lambda}_{\Phi \Psi}^2} 
~,
\nonumber\\
0 &=& P^{(f)}_T (\Omega_T) + {{\widetilde \lambda}_\Phi
{\widetilde \omega}^2  
- {\widetilde \lambda}_{\Phi \Psi} {\widetilde m}^2  
+ {\widetilde \lambda}_{\Phi \Psi} M_T^2  
- {\widetilde \lambda}_\Phi \Omega_T^2  
\over {\widetilde \lambda}_\Phi {\widetilde \lambda}_\Psi 
- {\widetilde \lambda}_{\Phi \Psi}^2} 
~.
\label{gapfinzztnew}
\end{eqnarray}
These can then be recombined to obtain 
the ``renormalized gap equations" 
\begin{eqnarray}
M_T^2 &=& {\widetilde m}^2 
+ {{\widetilde \lambda}_\Phi \over 2} \phi^2 
+ {{\widetilde \lambda}_\Phi \over 2} P^{(f)}_T(M_T)
+ {{\widetilde \lambda}_{\Phi \Psi} \over 2} P^{(f)}_T(\Omega_T)
~,
\nonumber\\
\Omega_T^2 &=& {\widetilde \omega}^2 
+ {{\widetilde \lambda}_{\Phi \Psi} \over 2} \phi^2 
+ {{\widetilde \lambda}_\Psi \over 2} P^{(f)}_T(\Omega_T)
+ {{\widetilde \lambda}_{\Phi \Psi} \over 2} P^{(f)}_T(M_T)
~.
\label{gapfinzzt}
\end{eqnarray}
In order to show that temperature independent renormalization
prescriptions also allow to renormalize the 
bubble effective potential, let me start by observing that
Eqs.(\ref{vsumazzt})-(\ref{voneTreg}) imply
\begin{eqnarray}
V_T &=&
{{m}^2 \over 2} \phi^2 
+ {{\lambda}_\Phi \over 24} \phi^4
+ Q_T^{(f)}[M_T] + Q_T^{(f)}[\Omega_T]
- {M_T^4 + \Omega_T^4 \over 4} I_2
- {M_T^2 + \Omega_T^2 \over 2} I_1
\nonumber\\
& & 
- {\lambda_\Phi  \over 8} 
\left( I_1 - M_T^2 I_2 + P_T^{(f)}[M_T] \right)^2
- {\lambda_\Psi \over 8} 
\left( I_1 - \Omega_T^2 I_2 + P_T^{(f)}[\Omega_T] \right)^2
\nonumber\\
& & 
- {\lambda_{\Phi \Psi}  \over 4} 
\left( I_1 - M_T^2 I_2 + P_T^{(f)}[M_T] \right)
\left( I_1 - \Omega_T^2 I_2 + P_T^{(f)}[\Omega_T] \right)
~. \label{vrensumai}
\end{eqnarray}
One can then use Eq.(\ref{gapdivzztnew}) to show that
\begin{eqnarray}
V_T \!\!\! &=& \!\!\!
{m^2 \over 2} \phi^2  +  {\lambda \over 24} \phi^4
- {M_T^4 + \Omega_T^4 \over 4}  I_2 
+ {M_T^2 + \Omega_T^2 \over 2}  I_1   
+ Q_T^{(f)}[M_T] + Q_T^{(f)}[\Omega_T] ~~~~~~~~~~~~~~~~~~~~~~~~~~~~~~~
\nonumber\\
& & 
\!\!\!\!\!\!\!\! \!\!\!- {\lambda_\Phi \over 2}
\left[ { \lambda_{\Phi \Psi} (\Omega_T^2 - \omega^2)
- \lambda_\Psi (M_T^2 - m^2)
+ (\lambda_\Phi \lambda_\Psi - \lambda_{\Phi \Psi}) {\phi^2 \over 2}
\over \lambda_\Phi \lambda_\Psi - \lambda_{\Phi \Psi}^2 } \right]^2
\nonumber\\
& & 
\!\!\!\!\!\!\!\! \!\!\!- {\lambda_\Psi \over 2}
\left[ { \lambda_{\Phi \Psi} (M_T^2 - m^2)
- \lambda_\Phi (\Omega_T^2 - \omega^2)
\over \lambda_\Phi \lambda_\Psi - \lambda_{\Phi \Psi}^2 } \right]^2
\nonumber\\
& & 
\!\!\!\!\!\!\!\! \!\!\!
- \lambda_{\Phi \Psi} \! { [ \lambda_{\Phi \Psi} 
(\Omega_T^2 \!\! - \!\! \omega^2)  \!
- \! \lambda_\Psi (M_T^2 \!\! - \!\! m^2)
\! + \! (\lambda_\Phi \lambda_\Psi \! 
- \! \lambda_{\Phi \Psi}) {\phi^2 \over 2}]
[ \lambda_{\Phi \Psi} (M_T^2 \!\! - \!\! m^2) \!
- \! \lambda_\Phi (\Omega_T^2 \!\! - \!\! \omega^2)]
\over [\lambda_\Phi \lambda_\Psi \! - \! \lambda_{\Phi \Psi}^2 ]^2 }
~, \label{vrensumazzttt}
\end{eqnarray}
which in terms of the renormalized parameters 
of Eqs.(\ref{renmlzza})-(\ref{renmlzze})
takes the form
\begin{eqnarray}
V_T \!\! \! &=& \!\! \! 
{{\widetilde m}^2 \over 2} \phi^2 +
{{\widetilde \lambda}_\Phi \over 24} \phi^4
+ {{\widetilde \lambda}_\Phi \! - \! {\lambda}_\Phi \over 12} \phi^4
+ Q_T^{(f)}[M_T] + Q_T^{(f)}[\Omega_T] ~~~~~~~~~~~~~~~~~~~~~~~~~~~~~~~~~~~~~
\nonumber\\
& & 
\! \! \!\!\!\!\!\!\!\!\!\!\! 
- {{\widetilde \lambda}_\Phi \over 2}
\left[ { {\widetilde \lambda}_{\Phi \Psi} (\Omega_T^2 - \omega^2)
- {\widetilde \lambda}_\Psi (M_T^2 - m^2)
+ ({\widetilde \lambda}_\Phi {\widetilde \lambda}_\Psi 
- {\widetilde \lambda}_{\Phi \Psi}) {\phi^2 \over 2}
\over {\widetilde \lambda}_\Phi {\widetilde \lambda}_\Psi 
- {\widetilde \lambda}_{\Phi \Psi}^2 } \right]^2
\nonumber\\
& & 
\! \! \!\!\!\!\!\!\!\!\!\!\! 
- {{\widetilde \lambda}_\Psi \over 2}
\left[ { {\widetilde \lambda}_{\Phi \Psi} (M_T^2 - m^2)
- {\widetilde \lambda}_\Phi (\Omega_T^2 - \omega^2)
\over {\widetilde \lambda}_\Phi {\widetilde \lambda}_\Psi 
- {\widetilde \lambda}_{\Phi \Psi}^2 } \right]^2
\nonumber\\
& & 
\! \! \!\!\!\!\!\!\!\!\!\!\! 
- {\widetilde \lambda}_{\Phi \Psi} \!
{ [{\widetilde \lambda}_{\Phi \Psi} 
(\Omega_T^2  \! \! -  \! \! \omega^2) \!
- {\widetilde \lambda}_\Psi (M_T^2 \!\! - \!\! m^2)
\! + \! ({\widetilde \lambda}_\Phi {\widetilde \lambda}_\Psi 
\! - \! {\widetilde \lambda}_{\Phi \Psi}) {\phi^2 \over 2}]
[ {\widetilde \lambda}_{\Phi \Psi} (M_T^2  \! \! -  \! \! m^2) \!
- \! {\widetilde \lambda}_\Phi (\Omega_T^2 \! \! - \! \! \omega^2)]
\over {\widetilde \lambda}_\Phi {\widetilde \lambda}_\Psi 
- {\widetilde \lambda}_{\Phi \Psi}^2 }
~, \label{vrensumaii}
\end{eqnarray}
and can be finally reexpressed using
the renormalized gap equations (\ref{gapfinzzt})
as
\begin{eqnarray}
V_T \! \! &=& \! \! 
{{\widetilde m}^2 \over 2} \phi^2 +
{{\widetilde \lambda}_\Phi \over 24} \phi^4
+ {{\widetilde \lambda}_\Phi \! - \! {\lambda}_\Phi \over 12} \phi^4
+ Q_T^{(f)}[M_T]
+ Q_T^{(f)}[\Omega_T]
\nonumber\\
& & - {{\widetilde \lambda}_\Phi \over 8} 
(P_T^{(f)}[M_T])^2
- {{\widetilde \lambda}_\Psi \over 8} 
(P_T^{(f)}[\Omega_T])^2
- {{\widetilde \lambda}_{\Phi \Psi} \over 4} 
P_T^{(f)}[M_T]
P_T^{(f)}[\Omega_T]
~, \label{vrensumaiii}
\end{eqnarray}
Once again the $\Lambda$-dependence is confined to the term
$({\widetilde \lambda}_\Phi \! - \! {\lambda}_\Phi) \phi^4 /12$
which [as easily checked from Eqs.(\ref{renmlzza})-(\ref{renmlzze})]
has a well-defined and finite $\Lambda \rightarrow \infty$ limit, 
indicating the renormalizability of
the CJT bubble potential 
of the $Z_2 \times Z_2$ model at finite temperature.

\section{CLOSING REMARKS}
$~$

In this paper I have provided evidence in support
of the reliability 
of approximations based on the CJT
formalism in the context of multi-field (thermal) theories.
The aspects of the analysis which required
the development of new technical tools,
such as the simultaneous renormalization of the
gap equations and the triviality-related issues,
might encode more physical information
than I was able to uncover here; 
this should be investigated in the future.
It would also be interesting to check whether 
in multi-field theories it is possible to confirm
the agreement established within single-field theories between
the nonperturbative approximations used in Refs.\cite{comp1,comp2}
and the CJT bubble approximation.
Issues such as those related to the
interdependence of the
gap equations are not easily phrased
in some of the other nonperturbative approximations,
and this might
lead to complications.

Most importantly,
the apparent reliability 
of the CJT formalism 
for multi-field theories
can be exploited in the investigation of physical
problems, especially the possibility of
symmetry nonrestoration, for which, as discussed in the introduction,
the CJT formalism is ideally suited.
The analysis presented in this paper already provides 
results directly relevant for symmetry nonrestoration;
in particular, from Eqs.(\ref{pqexp}) and (\ref{gapfinzzt})
it follows that at high temperatures
\begin{eqnarray}
M_T^2 &\simeq& {\widetilde m}^2 
+ {{\widetilde \lambda}_\Phi \over 2} \phi^2 
+ {{\widetilde \lambda}_\Phi \over 2}
({T^2 \over 12} - {M_T T \over 4 \pi})
+ {{\widetilde \lambda}_{\Phi \Psi} \over 2}
({T^2 \over 12} - {\Omega_T T \over 4 \pi})
~,
\nonumber\\
\Omega_T^2 &\simeq& {\widetilde \omega}^2 
+ {{\widetilde \lambda}_{\Phi \Psi} \over 2} \phi^2 
+ {{\widetilde \lambda}_\Psi \over 2} 
({T^2 \over 12} - {\Omega_T T \over 4 \pi})
+ {{\widetilde \lambda}_{\Phi \Psi} \over 2} 
({T^2 \over 12} - {M_T T \over 4 \pi})
~,
\label{gaphight}
\end{eqnarray}
which are the equations used in the 
symmetry nonrestoration 
analysis of Ref.\cite{sissa2}.
In Ref.\cite{sissa2} these equations were taken as a starting point,
without a discussion of the issues related to ultraviolet divergences;
therefore, my analysis,
by providing an explicit renormalization procedure
leading to (\ref{gaphight}),
renders more robust those results.
On the other hand my investigation of the ultraviolet structure
raises some new issues for
symmetry nonrestoration. Firstly, 
it appears that the physical consistency of the
analysis requires that the theory be considered 
as an effective low-energy theory.
Moreover, the triviality-related issues encountered
in Sec.3,
indicate that it is necessary to reconsider
the conventional assumption that (\ref{cond})
be sufficient for stability.
It appears, in fact, necessary
to perform
a more careful stability analysis analogous to 
the one given in Ref.\cite{mosh},
in which
it was shown that, as a by-product of triviality,
in the simple $Z_2$ model (the $\lambda \phi^4$ model),
the condition ${\widetilde \lambda}_\Phi \! > \! 0$
[whose equivalent in the $Z_2 \times Z_2$ model
is Eq.(\ref{cond})] is insufficient for stability.
Further investigation of this issue, and its relevance
for the possibility of
symmetry nonrestoration,
is left for future work.

\vglue 0.6cm
\leftline{\Large {\bf Acknowledgements}}
\vglue 0.4cm
It is a pleasure to acknowledge conversations with R. Jackiw, 
O. Philipsen, S.-Y. Pi, and S. Sarkar.

\newpage
\baselineskip 12pt plus .5pt minus .5pt

\end{document}